\documentclass[preprint]{revtex4}

\newcommand{\dee}{{\rm d}}

\usepackage{subfig}

\usepackage[pdftex]{graphicx}
\usepackage{epstopdf}
\usepackage{bm}
\usepackage{psfrag}
\usepackage{color}
\usepackage{times}
\usepackage{units}

\newcommand{\eq}{\begin{equation}}
\newcommand{\eeq}{\end{equation}}
\newcommand{\eqa}{\begin{eqnarray}}
\newcommand{\eeqa}{\end{eqnarray}}

\begin{document}

\title{A Turing instability in the solid state: void lattices in irradiated metals}

\author{MW Noble}
\affiliation{Department of Materials, University of Oxford, OX1 3PH, UK}
\author{MR Tonks}
\affiliation{Department of Materials Science and Engineering, University of Florida, 549 Gale Lemerand Drive, Gainesville, FL 32611, USA}
\author{SP Fitzgerald}
\email{S.P.Fitzgerald@leeds.ac.uk}
\affiliation{Department of Applied Mathematics, University of Leeds, LS2 9JT, UK}

\begin{abstract}\noindent 
Turing (or double-diffusive) instabilities describe pattern formation in reaction-diffusion systems, and were proposed in 
1952 as a potential mechanism behind pattern formation in nature, such as leopard spots and zebra stripes. Because 
the mechanism requires the reacting species to have significantly different diffusion rates, only a few liquid phase 
chemical reaction systems exhibiting the phenomenon have been discovered. In solids the situation is markedly 
different, since species such as impurities or other defects typically have diffusivities $\propto\!\exp\left( -E/k_{\rm B}
T\right)$, where $E$ is the migration barrier and $T$ is the temperature. This often leads to diffusion rates differing by 
several orders of magnitude. Here we use a simple, minimal model to show that an important class of emergent 
patterns in solids, namely void superlattices in irradiated metals, could also be explained by the Turing mechanism. 
Analytical results are confirmed by phase field simulations. The 
model (Cahn-Hilliard equations for interstitial and vacancy concentrations, coupled by creation and annihilation terms) 
is generic, and the mechanism could also be responsible for the patterns and structure observed in many solid state 
systems.
   
\end{abstract}

\maketitle

\noindent {\it Introduction ---} 
Patterns formed by Turing instabilities \cite{turing1952} arise in reaction-diffusion systems due to the competition 
between diffusion and nonlinear reaction terms. Counterintuitively, a uniform solution for reactant 
concentrations (known as a {\it base state}), stable in the absence of diffusion, can become unstable to the emergence 
of patterns and ordering once diffusion is switched on. This runs counter to the standard picture of diffusion as a 
smoothing influence, and is interesting to study from a non-equilibrium physics point of view. Some time after Turing's 
original prediction, chemical systems were discovered that exhibited the effect, though they remain rare since the 
Turing model typically requires the reacting species to diffuse at significantly different rates -- unusual in liquid phase 
chemical systems \cite{cross}. In the solid state, however, different species' diffusion rates generically differ by many 
orders of magnitude, since they are usually governed by nonlinear Arrhenius escape rates $\propto\!\exp\left( -E/k_{\rm 
B}T\right)$, where the migration barrier $E$ can vary from fractions-of to several eV. We note that crowdion defects in 
body-centred-cubic (bcc) metals have migration barriers too low for the Arrhenius formula to apply, and their diffusion 
rates are linear in temperature \cite{fitzgerald2008peierls,swinburne2013}.  

An intriguing and technologically important example of solid state pattern formation is void and gas bubble superlattice 
formation in irradiated metals. First observed in the 1970s 
\cite{evans1971observations,johnson1978helium,sikka1972superlattice}, the voids generated by the agglomeration of 
the 
radiation-induced vacancies can form an ordered superlattice under certain conditions. This runs counter to
 the more intuitive picture of Ostwald ripening, where large voids grow at the expense of smaller ones. Also, noble gases formed in fission reactors (e.g. 
 Kr, Xe) generally have very low solubility in metals, and hence segregate to regions of high tensile strain. At grain 
 boundaries, this leads to embrittlement, and accelerated mechanical failure. Engineering a stable bubble lattice 
 (formed of voids filled with gas atoms) potentially offers a way to sequester this gas atoms safely away from grain boundaries and 
 extend the life of reactor materials \cite{harrison2017engineering}.
 Superlattices 
 are most often observed within a temperature window of 0.2-0.4 of the melting point \cite{robinson2017effect}, and often mimic 
 the lattice 
 symmetry of the underlying crystal, though with a spacing tens or hundreds of times larger; see \cite{ghoniem2001theory} for a thorough review. 
 These lattices form over minutes and hours, 
 meaning molecular dynamics simulations cannot hope to directly capture the processes at work. 

Various competing mechanisms for superlattice formation have been proposed, including elastic interactions
 between voids, isomorphic decomposition, phase instability, interstitial dislocation loop punching and anisotropic 
 interstitial diffusion 
\cite{krishan1982invited,tewary1972theory,woo1985theory,khachaturyan1974spatially,evans1985computer,dubinko1986theory}. 
 Here we propose an alternative mechanism, and argue that void lattices could emerge
  as a Turing instability, where diffusion itself destabilizes the uniform base states which solve the steady-state, 
  diffusionless equations of motion. Whilst some or all of the mechanisms above may play a role in the details 
  of the superlattice formation, we show all that is actually required is a region in which local vacancy and interstitial 
  concentration, generation, and annihilation rates satisfy a specific relation, and
   vacancy and interstitial diffusion rates that are sufficiently different. Ours is the simplest 
   possible model that can capture 
the diffusion of two reacting species, with like species tending to cluster. It is a gross idealizeation, and neglects many 
important features of real crystal systems, in particular the anisotropic nature of self-interstitial diffusion and the 
elastic interactions between species. Nevertheless, it is sufficient to predict the formation and lengthscale 
of ordered patterns, as we show below. Our purpose here is to present a minimal and general model, which may be 
applied to many different systems, rather than to focus on the details of specific 
materials. A systematic study dealing with particular metals and radiation conditions will be published 
elsewhere. 

In the next section, we apply Turing's linearized analysis to the pair of coupled equations governing the 
diffusing defects, and extract analytical conditions for the system to support a superlattice of a given wavenumber. 
We then perform fully non-linear phase field simulations to investigate the system behaviour at longer times, 
confirming that the superlattice wavenumber predicted by the linear analysis is indeed realised in the full system.

\noindent {\it The model ---} In what follows, $v(\bm{x},t)$ and $s(\bm{x},t)$ denote the concentrations of vacancies and 
self-interstitials respectively. A phase field model \cite{moelans2008,chen2002} for their evolution leads to Cahn-Hilliard 
equations \cite{cahn1958}, with additional terms corresponding to creation ($c$) and annihilation ($ -a  sv$, according to the law of mass action): 
\eqa
\dot s & = & D_s\nabla^2\left(\frac{\delta F[s,v]}{\delta s}\right)+c- a  sv;\nonumber\\
\dot v & = & D_v\nabla^2\left(\frac{\delta F[s,v]}{\delta v}\right)+c- a  sv
\label{eq:CH1}
\eeqa The terms in brackets are functional derivatives of the following simple double-well free energy $F[s,v]$ with respect to 
$s$ and $v$:
\eq
F= \int_V \left[ s^2 \left(1-s\right)^2 +\frac{\gamma_s}{2} |\nabla s|^2 + 
v^2 \left(1-v\right)^2 +\frac{\gamma_v}{2} |\nabla v|^2
\right]\dee V.\eeq
The quartic bulk free energy terms have minima when the concentrations $s$ and $v$ are 0 or 1, encouraging the 
formation of voids and clusters. The $D$s are the diffusivities, with $D_s\gg D_v$ in metals, and the $\gamma$s are 
proportional to the square of the effective interface size between solid and void/cluster regions. We stress that all these 
parameters take {\it effective} values. Since superlattice formation takes place on a timescale of hours, the underlying atomic 
processes will be averaged over many realizations. For example, the annihilation rate $a$ does not represent the probability 
of mutual annihilation when a vacancy and self-interstitial atom meet, but rather the fraction of defects which annihilate over 
a representative region in a representative time interval. 

The explicit form of the 
equations is 
\eqa
\dot s & = & D_s\nabla^2\left(2s(s-1)(2s-1)-\gamma_s\nabla^2s\right)+c- a  sv;\nonumber\\
\dot v & = & D_v\nabla^2\left(2v(v-1)(2v-1)-\gamma_v\nabla^2v\right)+c- a  vs.\label{eq:CH2}
\eeqa These equations conserve the number of defects during evolution (apart from the explicit creation 
and annihilation terms), in contrast with the 
coupled rate equation model \cite{bullough1975cascade} explored in ref.\cite{ghoniem2001theory}, which involves only two spatial derivatives. Note that the 
defects do not interact 
until they meet and react: this is not a Fokker-Planck model of diffusion in a position-dependent potential, but rather 
a reaction-diffusion one. 

\noindent {\it Analytical results ---} We now follow the analysis due to Turing, and linearize the system about a so-called base state $\bar s,\, \bar v$ 
which satisfies the static equations, Eqs.(\ref{eq:CH2}) with all spatial and temporal derivatives set to zero: 
\eq
s(\bm{x}) = \bar s + S(\bm{x});\;\; v(\bm{x}) = \bar v + V(\bm{x});\;\; c -  a  \bar s\bar v = 0.\label{eq:base}
\eeq This imposes a relation between the uniform base states and the creation and annihilation rates. Seeking 
solutions of the form $(S,V)\equiv \bm{S} = \bm{S_0} \exp\left[ \lambda t + i\bm{q}\cdot\bm{x}\right]$ leads to the 
eigenvalue equation $\lambda\bm{S_0} = \bm{A_q S_0}$, with 

\eq
\bm{A_q} = \left(\begin{array}{cc}
-D_s(q^2g_s+\gamma_sq^4)- a \bar v & - a \bar s \\
- a \bar v & -D_v(q^2g_v+\gamma_vq^4)- a \bar s  \\
\end{array}\right),
\eeq where $q=|\bm{q}| = (q_x^2+q_y^2)^{1/2}$ in 2D, $g_s = 2(6\bar s(\bar s - 1)+1),$ and $ g_v = 2(6\bar v(\bar v - 1)+1)$. The eigenvalues 
$\lambda$ are given by the two solutions to det$(\bm{A_q}-\lambda\bm{I})=0$. If both
 solutions for $\lambda(q)$ are negative, the solution decays in time, and hence the base state is stable to 
 perturbations of wavenumber $\bm{q}$. A Turing instability arises when a base state is stable for $D_s=0=D_v$ 
 (equivalently  $q=0$), but becomes unstable  when it is 
 perturbed by a certain wavenumber $\bm{q}$. The growing solution then leads to periodic patterns with wavenumber 
 $\bm{q}$.

When $q=0$, $\lambda = 0$ or $ -a(\bar s + \bar v),$
so for all base states, no unstable ($\lambda>0$) pattern-forming mode is possible without diffusion. When diffusion is 
switched on, one or both eigenvalues are pushed above zero  when either ${\rm tr}\bm{A_q}  > 0$ and $({\rm tr}
\bm{A_q})^2 - 4\,{\rm det}\bm{A_q} > 0$, or ${\rm tr}\bm{A_q}  < 0$ and ${\rm det}\bm{A_q} < 0.$
A sufficient (but not necessary) condition for the Turing instability is hence $
{\rm det}\bm{A_q} < 0.$ Assuming $\gamma_s = \gamma_v$, and working in units where $\gamma_s = \gamma_v=1$
leads to 

\begin{figure}

\includegraphics[width=0.5\textwidth]{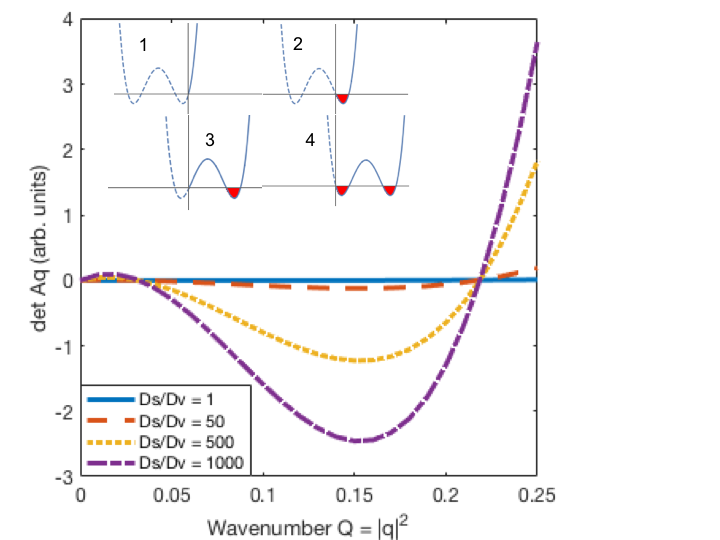}
\caption{Inset: Four scenarios leading to different regions of instability. Shaded regions show wavenumbers of possible patterns. Main plot: Deepening instability as $D_s/D_v\to$ 1, 50, 500, 1000. The minimum corresponds to the most negative 
eigenvalue, and hence the wavenumber with the fastest-growing instability. This is the wavenumber that the emergent 
pattern adopts, as confirmed by our numerical simulations.}\label{fig:inst}
\end{figure}

\begin{widetext}
\eq
{\rm det}\bm{A_q}  =  D_vD_s q^8 + D_vD_s(g_s + g_v)q^6
+(D_vD_s g_s g_v +  a(D_v \bar v+D_s\bar s))q^4 + a(D_v\bar v g_v + D_s\bar s g_s)q^2,\label{eq:detAq}
\eeq \end{widetext}a quartic in $q^2$, passing through $q^2=0$ (reflecting the 
conservation of 
vacancies and interstitials). Positive values of $q^2$ that lead to a negative value of det$\bm{A_q}$ correspond to 
a pattern with wavenumber $\bm{q}$. $q>2\pi$ is not physically interesting, since it corresponds to patterns of 
wavelength less than the interface width. Also, $q\to 0$ corresponds to complete decomposition into void and 
undefected crystal, thus the most predictive, and hence physically interesting, case is the third in Fig.\ref{fig:inst} (inset), 
where only a certain range of wavenumbers lead to instability.

Since the equation for the determinant is effectively a cubic, it can be solved 
analytically, and the value of the superlattice spacing $\Lambda$ can be extracted as a function of the input parameters. For case 3, this 
is given by $\Lambda = 2\pi/\sqrt{Q_*/2}$, where $Q_*$
is the largest root of ${\rm d}({\rm det}\bm{A_q})/{\rm d} Q = 0$ (see Fig. \ref{fig:inst}). 

$D_s\gg D_v$ means that the interstitials generated during a cascade diffuse away faster than 
the vacancies, typically 
leading to a ``halo'' of interstitials surrounding a region of high vacancy density. Setting $\bar v=0.25$ and $\bar s =
 0.01$ to reflect an example of this results in the third scenario described above. The determinant is shown in Fig.
 \ref{fig:inst} for
  several values of the diffusivity ratio ($a=1$ in this plot. The values of $a$ and $c$ are constrained by 
  Eq.(\ref{eq:base}). 
When $D_s = D_v$, the determinant barely dips below zero, but as the ratio $D_s/D_v$ increases up to the value 
of 1000 typical for bcc metals, the instability deepens. The minimum, most unstable, wavenumber $q$ for these 
parameters is approximately
 $\sqrt{0.15}$, corresponding to a pattern period $2\pi/(q/\sqrt 2)$ of about 23 times the interface width, or around 100 spacings of the underlying crystal lattice, if we take 
 the interface to be 4 crystal lattice spacings in width (again, this is an effective quantity, chosen to appropriately balance the bulk and interface terms in the 
 free energy, and need not correspond precisely to the size of the physical interface at the void surface). This is consistent with experimentally 
 observed void lattices. 
 
\begin{figure}
\begin{center}
\includegraphics[width=0.45\textwidth]{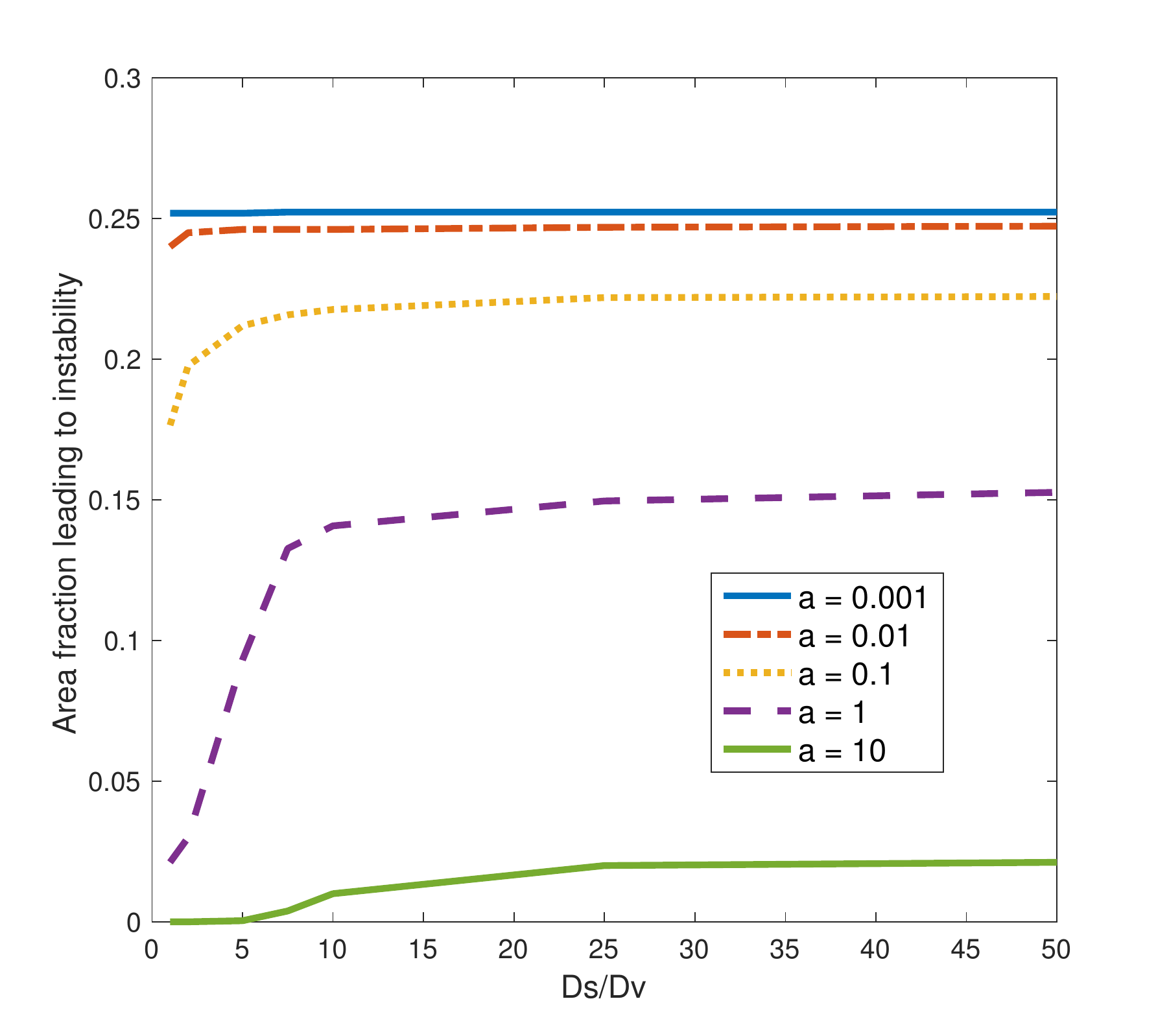}
\caption{Fractions of region $[0,0.5]\times [0,0.5]$ in $(\bar s,\bar v)$ parameter space leading to instability}\label{fig:areas}
\end{center}
\end{figure}

The above values for $\bar v$ and $\bar s$ represent a reasonable example, but in any irradiated crystal, different regions will have different 
values. The conditions for instability are not particularly restrictive, however. According to Descartes' rule of signs, a cubic has two positive roots (i.e. 
case 3 discussed above) when there are two sign changes between the successive terms in Eq. (\ref{eq:detAq}), and the discriminant is positive. Since the 
first coefficient is always positive, this means the last coefficient must be positive, and at least one of the second and third coefficients must be negative. 
Fig.\ref{fig:areas} shows the fraction of the region in parameter space defined by $(\bar s,\bar v)\in [0,0.5]\times [0,0.5]$ that satisfies these conditions, and hence 
supports a pattern-forming 
instability, as a function of the diffusivity ratio (we restricted the full $(u,v)\in [0,1]$ range to exclude unrealistic base states with $>50\%$ vacancies or interstitials). Several values of the annihilation 
parameter $a$ are shown. For each value of $a$, the 
unstable region reaches a plateau when the diffusivity ratio exceeds approximately 100. The lattice-forming region also grows as the effective 
annihilation parameter falls. For $a < 0.1$, around a quarter of the possible values for $\bar s$ and $\bar v$ lead to case 3 and hence an instability. This condition is again sufficient, 
but not necessary.

\begin{figure}
\subfloat[$D_s/D_v = $ 1 \label{fig:1to1}]{
        \includegraphics[width=0.15\textwidth]{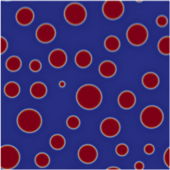}
       }
\subfloat[$D_s/D_v = $ 2 \label{fig:1to2}]{
        \includegraphics[width=0.15\textwidth]{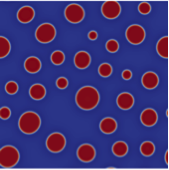}
       }
\subfloat[$D_s/D_v = $ 5 \label{fig:1to5}]{
        \includegraphics[width=0.15\textwidth]{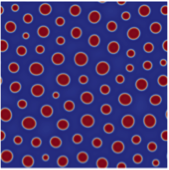}
       }\\
       \subfloat[$D_s/D_v = $ 10 \label{fig:1to10}]{
        \includegraphics[width=0.15\textwidth]{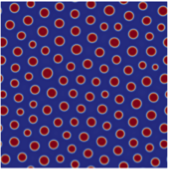}
       }
\subfloat[$D_s/D_v = $ 100 \label{fig:1to100}]{
        \includegraphics[width=0.15\textwidth]{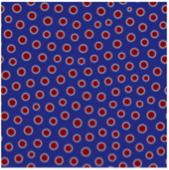}
       }
\subfloat[$D_s/D_v = $ 1000 \label{fig:1to1000}]{
        \includegraphics[width=0.15\textwidth]{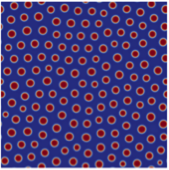}
       }
\caption{Phase field simulations of a 2D system governed by Eq.(\ref{eq:CH2}), for increasing values 
of the $D_s/D_v$ ratio. Ordering clearly emerges once $D_s/D_v\geq 5$, and the lattice spacing is insensitive 
to the ratio.}\label{fig:moose}
\end{figure}

\noindent {\it Phase field simulations ---} The Turing analysis is based on linearization, and it is reasonable to ask whether the 
patterns remain once the nonlinearity becomes important, and the nascent regions of high vacancy 
concentration grow into voids. We used the open source Multiphysics Object Oriented 
Simulation Environment 
\cite{tonks2012object,schwen2017rapid} to integrate Eqs.(\ref{eq:CH2}) numerically on a 2D domain, using the finite 
element method with implicit time integration, starting from an 
initial condition 
randomized about $\bar s =  0.25,\bar v=0.01$ with $a=0.5$. The results are shown in Fig. \ref{fig:moose}. Ordering is absent 
when $D_s/D_v<5$ and clearly emerges when  $D_s/D_v\geq 10$. The superlattice spacing is approximately 25 units, 
confirming that the system selects the fastest-growing unstable mode as predicted by the analytical model. The lattice is hexagonal, which is the 
expected symmetry that minimizes the free energy for a given wavenumber; the equivalent in three dimensions is 
body-centred-cubic (bcc) \cite{3DTuring}. 

\begin{figure}
\begin{center}
\includegraphics[width=0.4\textwidth]{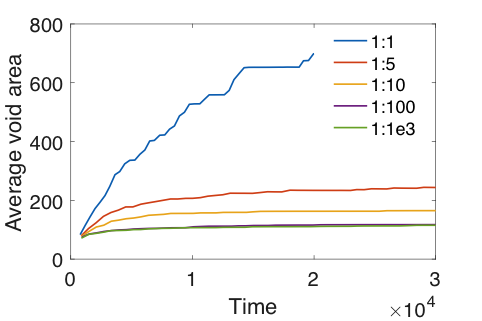}
\includegraphics[width=0.4\textwidth]{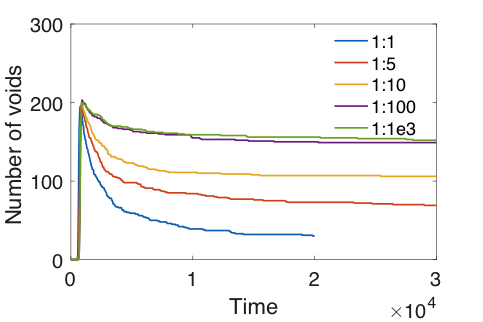}
\caption{Void area and number vs. time as $D_s/D_v\to$ 1, 50, 500, 1000. For equal diffusivities, the usual Ostwald ripening behaviour is evident. As their ratio grows, the stable Turing pattern emerges.}\label{fig:voids}
\end{center}
\end{figure}

Fig.\ref{fig:voids} shows the average void area 
and number of voids against time. Initially, voids nucleate in the regions where the fluctuating initial vacancy
concentration is high. For $D_s/D_v<5$, the standard picture of Ostwald ripening emerges, with large voids growing at 
the expense of smaller ones. As $D_s/D_v$ is increased however, the number of voids stabilizes, and an ordered 
lattice emerges, as is clear from Fig.\ref{fig:moose}. We simulated the system under a variety of different initial 
conditions, including pre-existing populations of voids of different sizes and distributions, and several different 
values for the creation term. In all cases with $D_s/D_v\gg 1$, we found a stable void lattice (see Supplementary 
Material). The voids do not nucleate in an ordered pattern, and the lattice begins to form after nucleation. Smaller 
voids on the lattice grow and larger voids shrink, and those not on lattice sites shrink until they disappear. Intriguingly, 
we observed diffusion-driven migration of established voids to lattice locations, consistent with experimental 
observations \cite{ghoniem2001theory}. This occurs in the absence of any advective term in the governing 
equations (\ref{eq:CH1}, \ref{eq:CH2}), and is purely due to the preferential diffusion of vacancies and interstitials so as to form the 
superlattice. This provides a mechanism for the fast migration of fairly large voids, which might intuitively be expected to be 
immobile.

\noindent {\it Discussion ---} We have shown that the simplest possible model for diffusing populations of vacancies and 
interstitials, subject to uniform creation and annihilation, supports void superlattice formation, even in the absence of 
refinements such as anisotropic interstitial diffusion and elastic interactions. The mechanism responsible for the 
ordering is the well-known Turing instability. This also offers a possible explanation for the observed 
temperature window for 
superlattice formation: the mechanism requires the diffusivities of the vacancies and interstitials to differ 
significantly. The ratio $D_s/D_v\propto \exp(-(E_{\rm mig}^{\rm int}-E_{\rm mig}^{\rm vac})/k_{\rm B}T)$, and since $E_{\rm mig}^{\rm int}<E_{\rm mig}^{\rm vac}$, it 
decreases at high temperature. At low temperatures, the vacancy diffusion rate is simply too slow for sufficient 
vacancies to cluster and form voids on experimental timescales. 

This simple model is sufficient to qualitatively account for most of the phenomena observed in void lattice formation: the 
temperature 
window for formation, bcc superlattices appearing in bcc crystals, and hexagonal superlattices in hexagonal 
crystals (where diffusion within the basal plane is sufficiently faster than diffusion normal to it to 
make 
the superlattices effectively 2D \cite{mazey1986bubble}). Our model cannot predict fcc lattices (which have more than one inherent lengthscale). 
We have also observed the unexpected purely diffusion-driven migration of 
established voids to superlattice sites. The lineararized Turing analysis predicts analytically the superlattice parameter in excellent agreement with fully 
nonlinear phase field simulations, even when the simulations are initialized with a pre-existing population of randomly distributed voids. The 
remarkable robustness of stable superlattice formation, together with the simple and general nature of the model, suggests that Turing 
instabilities and their associated patterns could be generic in many solid state systems, where widely differing diffusivities of different species are ubiquitous.

{\it Acknowledgments --- } SPF acknowledges useful discussions with Dr P. Edmondson and Prof S. Donnelly, and financial support from the UK EPRSC under grant number EP/R005974/1.


\section{Supplementary material}

In order to investigate the robustness of the pattern formation, we investigate the impact of various parameters on the pattern formation. In each of these analyses, we generate an initial set of randomly distributed voids and then let them evolve over time. The interstitial concentration throughout the domain is initialized at a value of $\bar s = 0.007$.

\newpage 

\noindent {\it (I) Impact of the initial average vacancy concentration on pattern formation---} Three simulations were conducted, starting with 90, 120, and 150 voids. The initial void radius was 5.8. This results in an initial average vacancy concentration of 0.137, 0.181, 0.225, respectively. In each case, a stable lattice of voids formed in the material. The production term in each was 0.00125.

\begin{figure}[h]
\subfloat[90, initial]{
        \includegraphics[width=0.25\textwidth]{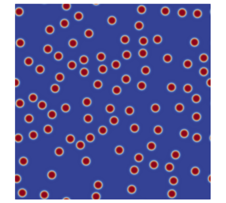}
       }
\subfloat[120, initial]{
        \includegraphics[width=0.25\textwidth]{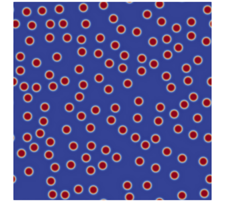}
       }
       \subfloat[150, initial]{
        \includegraphics[width=0.25\textwidth]{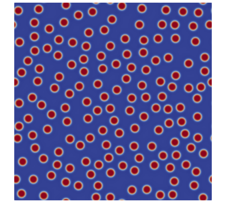}
       }\\
       \subfloat[90, final]{
        \includegraphics[width=0.25\textwidth]{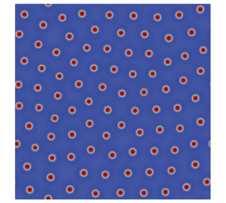}
       }
       \subfloat[120, final]{
        \includegraphics[width=0.25\textwidth]{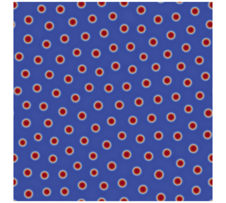}
       }
       \subfloat[150, final]{
        \includegraphics[width=0.25\textwidth]{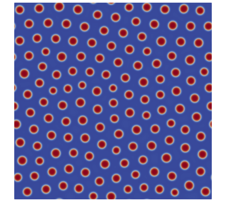}
       }\\
              \subfloat[Number of voids vs. time]{
        \includegraphics[width=0.32\textwidth]{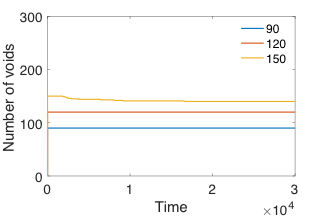}
       }
              \subfloat[Average void size vs. time]{
        \includegraphics[width=0.32\textwidth]{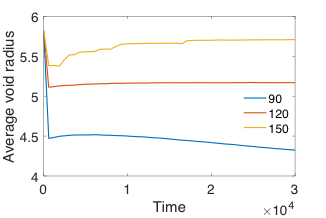}
       }
              \subfloat[Spread in void size vs. time (st. dev./mean)]{
        \includegraphics[width=0.32\textwidth]{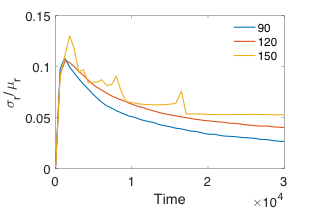}
       }
\caption{(I) Impact of the initial average vacancy concentration}\label{fig:moose1}
\end{figure}

\newpage

\noindent {\it (II) Impact of the magnitude of the source term---} Four simulations were conducted with different values for the defect production term, each starting with the same 90 voids. The four production terms were 0.0005, 0.00075, 0.0001, and 0.00125. 

\begin{figure}[h]
\subfloat[a = 0.0005]{
        \includegraphics[width=0.25\textwidth]{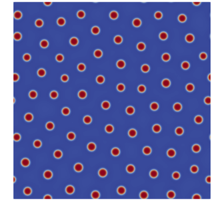}
       }
\subfloat[a = 0.00075]{
        \includegraphics[width=0.25\textwidth]{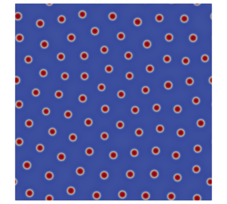}
       }
       \subfloat[a = 0.0001]{
        \includegraphics[width=0.25\textwidth]{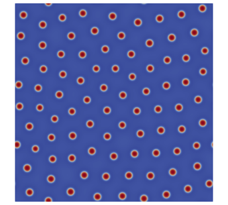}
       }
       \subfloat[a = 0.00125]{
        \includegraphics[width=0.25\textwidth]{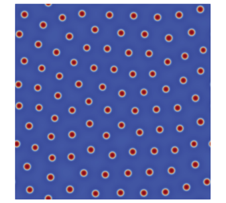}
       }\\
               \subfloat[Number of voids vs. time]{
        \includegraphics[width=0.32\textwidth]{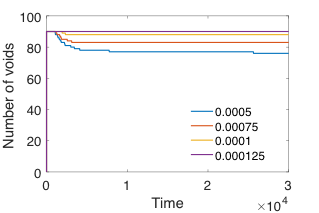}
       }
              \subfloat[Average void size vs. time]{
        \includegraphics[width=0.32\textwidth]{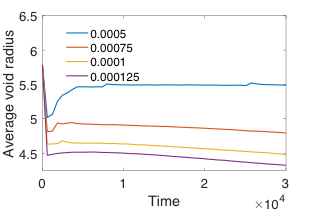}
       }
              \subfloat[Spread in void size vs. time (st. dev./mean)]{
        \includegraphics[width=0.32\textwidth]{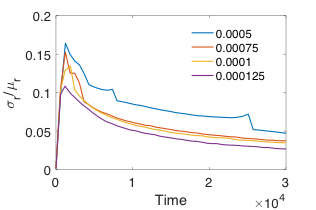}
       }
\caption{(II) Impact of the magnitude of the source term}\label{fig:moose2}
\end{figure}

\newpage

\noindent {\it (III) Impact of Variation in the Initial Void Size---}
In the previous simulations, each void started with the same size. Now, we compare the impact of randomly varying the initial size of the voids. In each case, the average void size is 5.8 and we start with 120 voids. We run one simulation with no variation, one in which the void size uniformly varies by $\pm$20\% of the void radius, and one which varies by $\pm$40\% of the void radius.

\begin{figure}[h]
\subfloat[No variation, initial]{
        \includegraphics[width=0.25\textwidth]{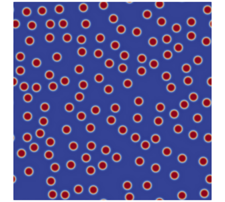}
       }
\subfloat[$\pm$20\%, initial]{
        \includegraphics[width=0.25\textwidth]{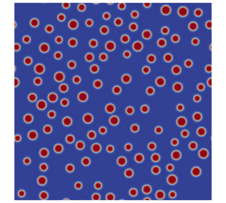}
       }
       \subfloat[$\pm$40\%, initial]{
        \includegraphics[width=0.25\textwidth]{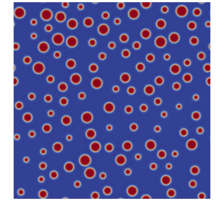}
       }\\
       \subfloat[No variation, final]{
        \includegraphics[width=0.25\textwidth]{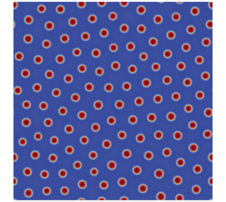}
       }
       \subfloat[$\pm$20\%, final]{
        \includegraphics[width=0.25\textwidth]{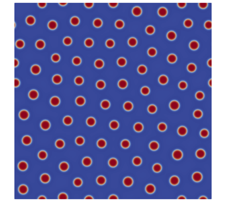}
       }
       \subfloat[$\pm$40\%, final]{
        \includegraphics[width=0.25\textwidth]{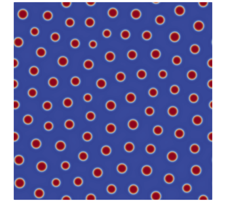}
       }\\
              \subfloat[Number of voids vs. time]{
        \includegraphics[width=0.32\textwidth]{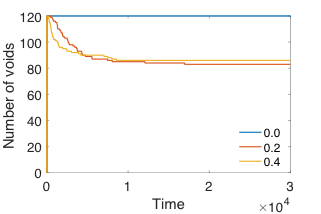}
       }
              \subfloat[Average void size vs. time]{
        \includegraphics[width=0.32\textwidth]{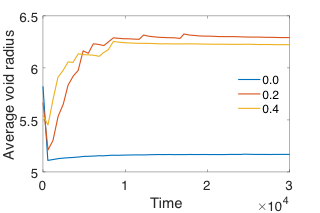}
       }
              \subfloat[Spread in void size vs. time (st. dev./mean)]{
        \includegraphics[width=0.32\textwidth]{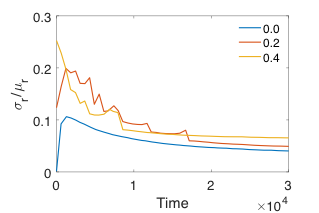}
       }
\caption{(III) Impact of variation in the initial void size}\label{fig:moose3}
\end{figure}

\end{document}